\documentclass{aa}  

\usepackage{graphicx}
\usepackage{txfonts}
\DeclareUnicodeCharacter{2212}{-}

\begin{document}

   \title{Infrared Spectroscopy of free-floating planet candidates in Upper Scorpius and Ophiuchus}

  \author{ H. Bouy \inst{1}
          \and M. Tamura \inst{2,3}
          \and D. Barrado \inst{4}
          \and K. Motohara \inst{3}
          \and N. Castro Rodríguez\inst{5}
          \and N. Miret-Roig\inst{1,6}
          \and M. Konishi\inst{2}
          \and S. Koyama\inst{7}
          \and H. Takahashi\inst{8}        
          \and N. Hu\'elamo\inst{4}
          \and E. Bertin \inst{9}
    	  \and J. Olivares\inst{10}
          \and L.~M. Sarro\inst{10}
          \and A. Berihuete\inst{11}
          \and J.-C. Cuillandre \inst{12}
          \and P.A.B. Galli\inst{13}
          \and Y. Yoshii\inst{7,14}
          \and T. Miyata\inst{7}
}

   \institute{Laboratoire d'astrophysique de Bordeaux, Univ. Bordeaux, CNRS, B18N, allée Geoffroy Saint-Hilaire, 33615 Pessac, France.
         \email{herve.bouy@u-bordeaux.fr}
         \and
         Department of Astronomy, Graduate School of Science, The University of Tokyo, Tokyo, Japan
         \and 
         National Astronomical Observatory of Japan,  Tokyo, Japan
         \and 
         Centro de Astrobiología (CSIC-INTA), Depto. de Astrofísica, ESAC Campus, Camino Bajo del Castillo s/n, 28692, Villanueva de la Cañada, Madrid, Spain
         \and
         Grantecan S. A., Centro de Astrofísica de La Palma, Cuesta de San José, 38712 Breña Baja, La Palma, Spain
         \and 
         University of Vienna, Department of Astrophysics, T\"urkenschanzstra\ss e 17, 1180 Wien, Austria
         \and 
         Institute of Astronomy, Faculty of Science, the University of Tokyo, Osawa 2-21-1, Miaka, Tokyo 181-0015, Japan
         \and
         Kiso Observatory, Institute of Astronomy, Faculty of Science, the University of Tokyo, Mitake 10762-30,
  Kiso-machi, Kiso-gun, Nagano 397-0101, Japan
         \and 
         Canada − France − Hawaii Telescope Corporation, 65-1238 Mamalahoa Highway, Kamuela, HI 96743, USA
         \and 
         Depto. de Inteligencia Artificial , UNED, Juan del Rosal, 16, 28040 Madrid, Spain 
         \and
		Depto. Estad\'istica e Investigaci\'on Operativa. Universidad de C\'adiz, Avda. Rep\'ublica Saharaui s/n, 11510 Puerto Real, C\'adiz, Spain
		\and 
		AIM, CEA, CNRS, Université Paris-Saclay, Université de Paris, F-91191 Gif-sur-Yvette, France 
		\and 
		Núcleo de Astrofísica Teórica, Universidade Cidade de São Paulo, R. Galvão Bueno 868, Liberdade, 01506-000, São Paulo, SP, Brazil
		\and
		Steward Observatory, University of Arizona, 933 North Cherry Avenue, Rm. N204 Tucson, AZ 85721-0065, USA
        }

   \date{Received ; accepted }

 
  \abstract
   {A rich population of low-mass brown dwarfs and isolated planetary mass objects has been reported recently in the Upper Scorpius and Ophiuchus star forming complex.}
   {We investigate the membership, nature and properties of 17 of these isolated planetary mass candidates using low-resolution near-infrared spectra.}
   {We investigate the membership by looking for evidences of youth using four diagnostics: the slope of the continuum between the J and Ks band, the H$_{\rm cont}$ and TLI-g gravity sensitive indices, and by comparing the spectra to
young and field (old) M and L-dwarf standards.}
   {All the targets but one are confirmed as young ultracool objects, with spectral types between L0 and L6 and masses in the range 0.004--0.013~M$_{\odot}$ according to evolutionary models. The status of the last target is unclear at this point.}
   {Only one possible contaminant has been identified among the 17 targets, suggesting that the contamination level of the original sample must be low ($\lesssim$6\%).}

   \keywords{Stars: brown dwarfs; Planets and satellites: formation }

   \maketitle
%

\section{Introduction}
Free-floating planets are planetary-mass objects that do not orbit a star, but roam the galaxy isolated. Apart from micro-lensing detections, only a few tens of directly imaged free-floating planet candidates are known to date \citep[for example among others ][ and references therein]{Tamura1998,Lucas2000,Zapatero2000,Luhman2005,Pena2012,Suarez2019,2018MNRAS.473.2020L,Luhman2020} and only a small fraction have been confirmed so far. Because of the degeneracy in the mass-luminosity relationship for these ultracool objects, it is indeed impossible to distinguish a low mass brown dwarf from a planetary mass object when the age and distance are unknown. This deadlock can be overcome by studying free-floating planets members of young associations where the age and distance are precisely known. In the context of the COSMIC-DANCE\footnote{\url{project-dance.com}} survey \citet{2022NatAs...6...89M} obtained deep optical and near-infrared photometry and measured accurate proper motions 5 mag beyond Gaia's limit in the nearby Upper Scorpius OB association (USco) and $\rho$-Ophiuchus (Oph) molecular clouds. They identified over 3\,500 members, including between 70 and 170 planetary mass objects, depending on the age assumed. This large number of planetary mass object candidates in a young association has important implications for the theories and models of star, brown dwarf and planet formation. In order to confirm this important discovery, we performed follow-up spectroscopic observations of 18 free-floating planet candidates to confirm their nature and membership to the association and validate \citet{2022NatAs...6...89M} analysis. In the following, we describe the observations and the processing of the data obtained at the Grantecan and Subaru telescopes. We discuss the membership to the association by looking for spectral features characteristic of young ultracool objects. Finally, we estimate the spectral types of the objects and derive effective temperature and mass estimates, as well as a contamination rate in \citet{2022NatAs...6...89M} sample.

\section{Observations}

\subsection{Targets}
\begin{figure}
   \centering
   \includegraphics[width=0.48\textwidth]{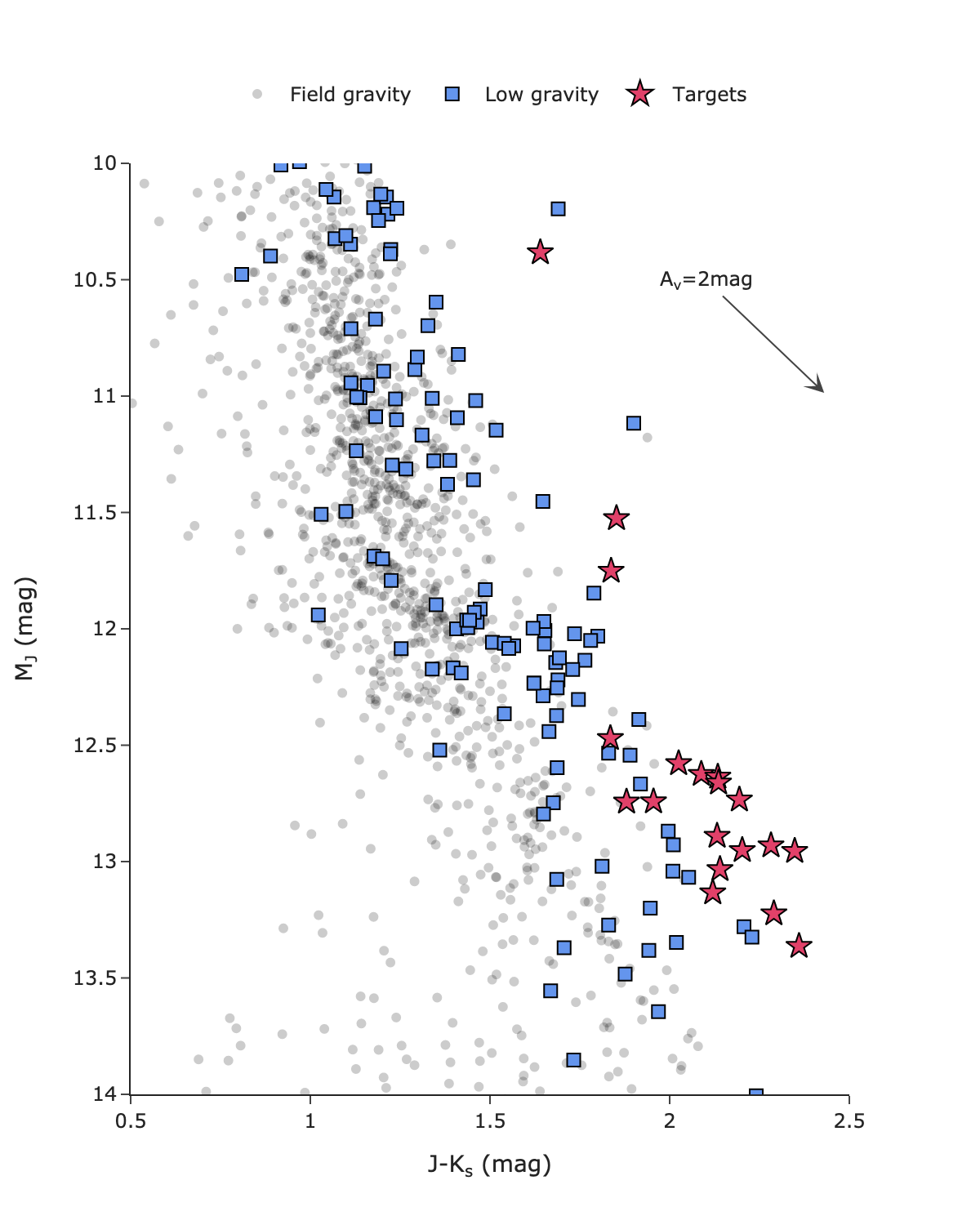}
      \caption{$(M_J,J-Ks)$ colour-magnitude diagram of our targets (red stars), with low-gravity (blue squares) and field gravity (grey dots) ultracool dwarfs from the literature \citep[][and references therein]{BurgasserSpexLib}. An arrow represents a $A_{\rm V}=2$~mag extinction vector.}
         \label{fig:JJKs}
\end{figure}

\begin{figure}
   \centering
   \includegraphics[width=0.45\textwidth]{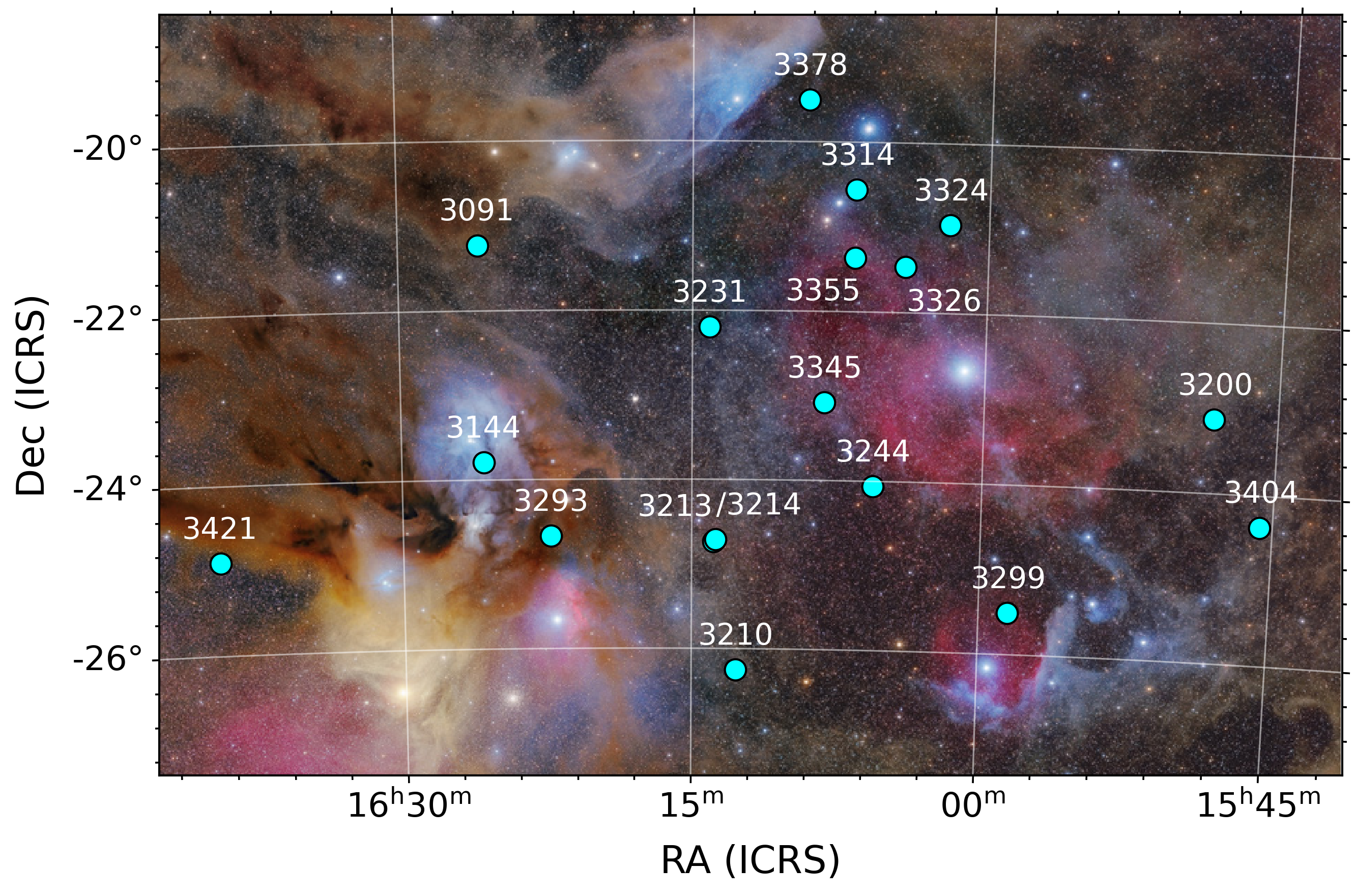}
      \caption{Position of our targets in Upper Scorpius and Ophiuchus. Background photograph credit: Mario Cogo (galaxlux.com)}
         \label{fig:map}
\end{figure}

We selected 18 targets within \citet{2022NatAs...6...89M} sample, after discarding objects already observed spectroscopically in the literature \cite[e.g.][]{2018MNRAS.473.2020L,Luhman2022}. A total of 18 objects were randomly selected in the range between $17.3<J<19.2$~mag, corresponding to masses between $7\lesssim $M$\lesssim10$M$_{\rm Jup}$ and effective temperatures between $1\,500\lesssim M\lesssim 1\,900$~K according to \citet{Baraffe+2002} evolutionary models for an age of 5~Myr and a distance of 145~pc. One brighter target (DANCe J16064553-2121595 = 3355, $J=16.19$~mag) was added during the course of the observations as clouds were degrading the sensitivity and preventing us to observe our original targets. Figure~\ref{fig:JJKs} shows a $(M_J,J-Ks)$ color-magnitude diagram of the sample and Fig.~\ref{fig:map} shows the location of the targets in Upper Scorpius and Ophiuchus. Two sources (3213 and 3214) are separated by only 120\arcsec. 

\begin{table*}
\centering
\caption{Targets observed
\label{tab_targets}}
\scriptsize
\begin{tabular}{cccccccccccc}
\hline
  Name & ID\tablefootmark{a} &   RA &   Dec &   $A_v$ max. &   i   & z &  y &   J  &   H  &   Ks  & Exp. Time \\
     & & (J2000) & (J2000) & mag     & mag  & mag   & mag  & mag & mag  & mag   & \texttt{NDIT$\times$DIT} s \\
\hline
\multicolumn{12}{c}{SWIMS} \\
\hline
DANCe J16135217-2443562  &  3213 & 16:13:52.17 & -24:43:56.2 & 0.82 & 23.76$\pm$0.13 & 22.03$\pm$0.05 & 21.52$\pm$0.06 & 18.84$\pm$0.06 & 17.70$\pm$0.05 & 16.70$\pm$0.05 & 40$\times$300\\				   
DANCe J16134589-2442310  &  3214 & 16:13:45.89 & -24:42:31.0 & 0.82 & 23.72$\pm$0.20 & 22.05$\pm$0.06 & 21.37$\pm$0.05 & 19.17$\pm$0.08 & 17.73$\pm$0.05 & 16.81$\pm$0.05 & 36$\times$300\\				   
DANCe J16081299-2304316  &  3345 & 16:08:12.99 & -23:04:31.6 & 0.52 & 23.69$\pm$0.08 & 22.01$\pm$0.05 & 22.15$\pm$0.08 & 18.94$\pm$0.05 & 17.80$\pm$0.05 & 16.82$\pm$0.05 & 24$\times$300\\				   
DANCe J16064553-2121595  &  3355 & 16:06:45.53 & -21:21:59.5 & 0.49 & 21.03$\pm$0.05 & 19.11$\pm$0.05 & 18.40$\pm$0.05 & 16.19$\pm$0.05 & 15.31$\pm$0.05 & 14.55$\pm$0.05 & 8$\times$300\\				   
DANCe J16091010-1930376  &  3378 & 16:09:10.10 & -19:30:37.6 & 0.60 & 24.09$\pm$0.11 & 22.19$\pm$0.05 & 21.58$\pm$0.05 & 19.03$\pm$0.09 & 17.81$\pm$0.05 & 16.74$\pm$0.05 & 22$\times$300\\				   
DANCe J15454024-2422072  &  3404 & 15:45:40.24 & -24:22:07.2 & 0.74 & 23.41$\pm$0.10 & 21.99$\pm$0.07 &$\cdots$ & 18.55$\pm$0.06 & 17.55$\pm$0.05 & 16.67$\pm$0.05 & 22$\times$300\\		
\hline
\multicolumn{12}{c}{EMIR} \\
\hline  
DANCe J16221885-2440029  &  3293 & 16:22:18.85 & -24:40:02.9 & 2.99 & 22.20$\pm$0.05 & 20.42$\pm$0.05 & 19.70$\pm$0.05 & 17.33$\pm$0.05 & 16.25$\pm$0.05 & 15.48$\pm$0.05 & 16$\times$200\\				   
DANCe J16064467-2033428  &  3314 & 16:06:44.67 & -20:33:42.8 & 0.77 & 22.29$\pm$0.05 & 20.45$\pm$0.05 & 19.85$\pm$0.05 & 17.56$\pm$0.05 & 16.50$\pm$0.05 & 15.72$\pm$0.05 & 28$\times$200\\				   
DANCe J16140756-2211524  &  3231 & 16:14:07.56 & -22:11:52.4 & 0.63 & 23.02$\pm$0.05 & 21.09$\pm$0.05 & 20.78$\pm$0.05 & 18.28$\pm$0.05 & 17.28$\pm$0.05 & 16.44$\pm$0.05 & 24$\times$200\\				   
DANCe J16255679-2113354  &  3091 & 16:25:56.79 & -21:13:35.4 & 1.81 & 23.53$\pm$0.11 & 21.76$\pm$0.05 & 21.19$\pm$0.06 & 18.39$\pm$0.08 & 17.33$\pm$0.05 & 16.36$\pm$0.05 & 28$\times$200\\				   
DANCe J16123953-2614510  &  3210 & 16:12:39.53 & -26:14:51.0 & 0.88 & 23.05$\pm$0.14 &$\cdots$& 20.24$\pm$0.05 & 18.43$\pm$0.06 & 17.31$\pm$0.05 & 16.34$\pm$0.05 & 24$\times$200\\			   
DANCe J16053908-2403330  &  3244 & 16:05:39.08 & -24:03:33.0 & 0.36 & 23.14$\pm$0.07 & 21.40$\pm$0.05 & 20.87$\pm$0.07 & 18.47$\pm$0.05 & 17.33$\pm$0.05 & 16.33$\pm$0.05 & 32$\times$200\\				   
DANCe J16020000-2057342  &  3324 & 16:02:00.00 & -20:57:34.2 & 0.66 & 23.15$\pm$0.05 & 21.66$\pm$0.05 & 21.11$\pm$0.05 & 18.54$\pm$0.05 & 17.50$\pm$0.05 & 16.35$\pm$0.05 & 32$\times$200\\				   
DANCe J16393029-2454135  &  3421 & 16:39:30.29 & -24:54:13.5 & 0.69 & 23.36$\pm$0.07 & 21.69$\pm$0.06 & 20.82$\pm$0.05 & 18.55$\pm$0.05 & 17.44$\pm$0.05 & 16.59$\pm$0.05 & 32$\times$200\\				   
DANCe J16041234-2127472  &  3326 & 16:04:12.34 & -21:27:47.2 & 0.52 & 23.69$\pm$0.07 & 21.91$\pm$0.05 & 21.53$\pm$0.05 & 18.70$\pm$0.05 & 17.43$\pm$0.05 & 16.57$\pm$0.05 & 32$\times$200\\				   
DANCe J16254583-2347372  &  3144 & 16:25:45.83 & -23:47:37.2 & 3.21 & 23.64$\pm$0.14 & 21.93$\pm$0.06 & 21.42$\pm$0.05 & 18.74$\pm$0.05 & 17.49$\pm$0.05 & 16.46$\pm$0.05 & 46$\times$200\\				   
DANCe J15481655-2307430  &  3200 & 15:48:16.55 & -23:07:43.0 & 0.77 & 23.65$\pm$0.12 &$\cdots$&$\cdots$& 18.76$\pm$0.09 & 17.50$\pm$0.05 & 16.56$\pm$0.05 & 32$\times$200\\			   
DANCe J15582895-2530319  &  3299 & 15:58:28.95 & -25:30:31.9 & 0.36 & 23.70$\pm$0.12 & 22.01$\pm$0.05 & 21.41$\pm$0.06 & 18.76$\pm$0.06 & 17.40$\pm$0.05 & 16.41$\pm$0.05 & 32$\times$200\\                            
\hline
\end{tabular}
\tablefoot{$A_v$ max. corresponds to the integrated line-of-sight interstellar reddening to 250~pc in the direction of the source as computed by \citet{Green2019}. {\bf The near-infrared JHKs photometry comes in the 2MASS system.}\\
\tablefoottext{a}{From \citet{2022NatAs...6...89M}}}
\end{table*}

\subsection{SWIMS at Subaru}
A total of six objects were observed with the \emph{Simultaneous-color Wide-field Infrared Multi-object Spectrograph} \citep[SWIMS,][]{10.1117/12.2054861, Motohara2016, Konishi2018, Konishi2020} mounted on the Subaru Telescope in May 2021 (Program S21A-047, PI: M. Tamura). SWIMS was used in long-slit mode with its simultaneous \emph{zJ} ($700<R<1200$) and \emph{HK} (600<R<1000) grisms. The 300~s individual exposures were acquired following a standard ABBA procedure to efficiently remove the sky emission.  
The seeing was generally very good (between 0\farcs3 and 0\farcs6) but clouds were hindering the observations at times. A slit of 0\farcs5 or 0\farcs8 was used depending on the seeing. Table~\ref{tab_targets} gives the list of targets observed with SWIMS and the corresponding number of exposures and individual exposure times. Three B stars of the Upper Scorpius associations were observed (HIP81145, HIP82133 and HIP78702) to be used as telluric standards.

The raw data were processed following standard procedures for infrared spectroscopy using a combination of custom made Python code using the \emph{astropy} and \emph{specutils} libraries \citep{astropy2018,specutils} and IRAF/PyRAF's \emph{apall} package for the spectra extraction and telluric correction. The closest-in-time B stars spectrum was used to remove the telluric contamination in each of our targets spectra. Because the B stars belong to Upper Scorpius as well, the typical airmass differences with the target were less than 0.2$\sim$0.3.

\subsection{EMIR at GTC}
A total of 13 objects were observed with \emph{Espectrógrafo Multiobjeto Infra-Rojo} \citep[EMIR,][]{2014SPIE.9147E..0UG} mounted on the Grantecan telescope (Program GTC2-21A, PI: D. Barrado) in May 2021. EMIR was used in long-slit mode with its \emph{HK} grism and a slit of 1\farcs2 chosen to match the ambient seeing during the observations, and leading to an effective resolution of $R\sim500$. The weather was mostly clear during the observations. Table~\ref{tab_targets} gives the list of targets observed with EMIR and the corresponding number of exposures and exposure times. The 200~s individual exposures were acquired following a standard ABBA procedure to efficiently remove the sky emission.  

The data were reduced using \emph{RedEmIR}, a new GTC pipeline written in Python. \emph{RedEmIR} eliminates the contribution of the sky background in the near infrared using the consecutive A-B pairs. The sky-subtracted images are subsequently flat-fielded, calibrated in wavelength and average combined to obtain the final spectrum. The telluric correction is achieved using a customized version of Xtellcor \citep{2003PASP..115..389V} adjusted to the atmospheric conditions of the La Palma observatory \citep{2009ApJ...694.1379R}. The spectra are then divided by the  spectrum of an A0 star spectrum obtained after or before the targets to remove telluric contamination.

\section{Evidences of youth}

Young ultra-cool dwarfs such as the ones targeted in the present study have not yet contracted into their final configurations and their gravity is significantly lower than their older field counterparts. 
At the resolution of our spectroscopic observations, gravity will affect mostly three features:

\begin{itemize} 
\item $J$ to $Ks$ continuum slope: a $J-Ks$ color redder than for field brown dwarfs has been systematically reported for young brown dwarfs \citep[e.g.][]{Kirkpatrick2008,Delorme2017b} as well as some young planets and brown dwarf companions to stars \citep[e.g.][ and Fig.~\ref{fig:JJKs}]{Barman2011, Delorme2017a}. The lower gravity leads to more clouds in the upper layers of the atmosphere. These reduce the amount of emergent flux at shorter wavelengths and lead to fainter absolute $J$-band magnitudes and redder $J-Ks$ colors. Additionally, the lower density associated to a lower gravity results in reduced collision-induced $H_2$ and FeH absorption which in turn leads to less suppression of the K-band flux and therefore a redder $J-Ks$ color \citep[][]{Mohanty2007, Faherty2013}.
\item triangular H-band continuum: only part of the $H$-band flux is affected by this reduced collision-induced FeH and $H_2$ absorption, producing an $H$-band continuum with a characteristic triangular shape \citep{Lucas2001}, which can be quantified and measured using the H$_{\rm cont}$ index proposed by \citet{Allers2013}. It is insensitive to reddening but the presence of dust in the upper layer of the atmosphere can mimic the effect of low-gravity on the H-band shape. For this reason, \citet{Allers2013} recommends to complement the H$_{\rm cont}$ index with other diagnostics to test the youth of ultracool objects.
\item gravity-sensitive absorption lines: collision induced pressure broadening depends on both the temperature and gravity (through the density) in the ultracool dwarf atmosphere. For a given effective temperature, an older ultracool dwarf with a higher gravity will therefore have more prominent absorption lines than a younger (lower gravity) counterpart \citep{Martin1996,Gorlova2003,Allers2013}. At the resolution of our observations, the 1.244 and  1.253~$\mu$m \ion{K}{I} lines are the most gravity sensitive lines detectable in the SWIMS spectra. Unfortunately, the relatively low signal-to-noise ratio of our SWIMS spectra in the J-band results in large uncertainties and inconclusive values of the KI$_{\rm J}$ index of \citet{Allers2013}.
\item TLI-g gravity sensitive index: taking advantage of the growing number of near-infrared spectra available in the literature and in various databases, \citet{Almendros2022} recently used machine learning techniques to define a new gravity sensitive index. Their TLI-g index is designed to separate young objects from older field objects with a performance superior to other indices from the literature. It seems in particular less sensitive to the presence of dust in the upper layer of the atmosphere, but is however sensitive to extinction. 
\end{itemize}

In the following, we therefore discuss the $J-Ks$ color, H$_{\rm cont}$ and TLI-g indices of our targets and compare their spectra to young and field M and L-dwarf standards.

\subsection{$J-Ks$ colours}
Figure~\ref{fig:JJKs} shows that all our targets have $J-Ks$ colors redder than older field counterparts from the literature and similar to known young low-gravity ultracool dwarfs. Both multiplicity and extinction can shift objects in this diagram and mimic the effect of youth. The presence of an unresolved companion can indeed shift the  position mostly vertically in a colour-magnitude diagram and by at most 0.75~mag\footnote{For an equal mass binary the individual fluxes are half the combined flux and the individual magnitudes are $2.5 \log_{10}(2)=0.75$~mag fainter.}. While we cannot rule out the presence of unresolved companions, Figure~\ref{fig:JJKs} shows that our targets remain redder than most field L-dwarfs even when adding 0.75~mag to their luminosity. Any unresolved pair would therefore be made of individual components redder than the older field sequence. Extinction is unlikely to have shifted the objects given that the cumulative line-of-sight reddening towards our objects is low in most cases (see Table~\ref{tab_targets}) and still places them on the redder low-gravity sequence even in the worst cases of 3293 and 3144. While it is not a conclusive proof, the very red $J-Ks$ color certainly adds to the list of evidences indicating youth and hence membership to USco or Oph.

\subsection{Comparison with young and old M and L standards}

To further assess the youth of our targets and at the same time derive their spectral type, we perform an empirical comparison with spectra of young and old ultracool objects from the literature. The comparison is made using \emph{The SpeX Prism Library Analysis Toolkit} \citep[SPLAT,][]{BurgasserSplat}. A number of spectral libraries of young ultracool dwarfs have been presented in the literature, and we chose to use the very-low gravity spectral standards included in SPLAT \citep[see Table~\ref{splat_stds} and][]{BurgasserSplat}. The field-gravity (older) standards chosen for the comparison are presented in Table~\ref{old_stds}. 

The degeneracy between extinction and spectral type can affect and compromise the comparison \citep[see e.g.][]{Luhman2017}. To partially lift this degeneracy and explore the effect of extinction on the results of our analysis, we perform the comparison after dereddening our spectra by the cumulative extinction in the line-of-sight of each target up to 250~pc reported in the 3D extinction map of \citet{Green2019}. Assuming that all our targets belong to USco, this should represent a worst case scenario in terms of reddening and provide an estimate of the lower limit on the spectral type. USco is indeed located at approximately 145~pc and Oph at 125~pc, although they possibly extend up to $\sim$200~pc \citep{Damiani2019}. The value of 250~pc was therefore chosen to be conservative and be sure to include the entire depth of the clouds associated to these two regions.

\emph{SPLAT} was first used to scale the instrumental fluxes to physical units using the target H-band photometry  reported in \citet{2022NatAs...6...89M} as reference. The closest match in each of the two libraries of standards is found using a standard $\chi^2$-minimization. Figures~\ref{fig:spt1} to \ref{fig:spt3} show the results, with the original spectra on the left and the dereddened spectra on the right. As expected the spectral types obtained using the high-gravity (older) field standards are systematically later than those obtained with the low-gravity (young) standards. 

Figures~\ref{fig:spt1} to \ref{fig:spt3} show that 3210, 3200, 3314, 3091, 3326, 3214, 3355, 3144, 3299, 3345 are clearly better matched by a young spectral standard than by an old spectral standard independently of the reddening.

Among the rest of the sample we can see that:
\begin{itemize}
\item 3378 and 3213 have K-band fluxes significantly higher than the best-match field standards which is clearly pointing towards a young age.
\item the best matches for 3244, 3421, 3231 and 3404 are obtained with a young standard but the $\chi^2$ difference is only marginal and inconclusive.
\item 3293 has a marginally better fit with an old standard but the $\chi^2$ difference is only marginal as well.
\item 3324 displays a very peculiar spectral energy distribution, that cannot be matched by any of our old or young standards. While we cannot rule out that the peculiar continuum emission is related to variability, we note that  3324 was observed during degraded ambient conditions and the discrepancy is probably due to telluric clouds absorption, as suggested by the discrepancy between the $H-K_{s}$=1.15~mag photometry reported by \citet{2022NatAs...6...89M} and the synthetic $H-K_{s}$=0.8~mag computed from the spectral data. This spectrum is therefore considered as dubious and discarded for the rest of the analysis.
\end{itemize}

\subsection{Sharp H-band continuum}
 As mentioned earlier the shape of the H-band continuum varies from a typical triangular shape at young ages to a flatter continuum at more advanced ages as gravity increases. The H$_{\rm cont}$ index defined by \citet{Allers2013} is commonly used to quantify the sharpness of the H-band continuum and look for evidence of youth. 
 
\begin{figure*}
   \centering
   \includegraphics[width=0.95\textwidth]{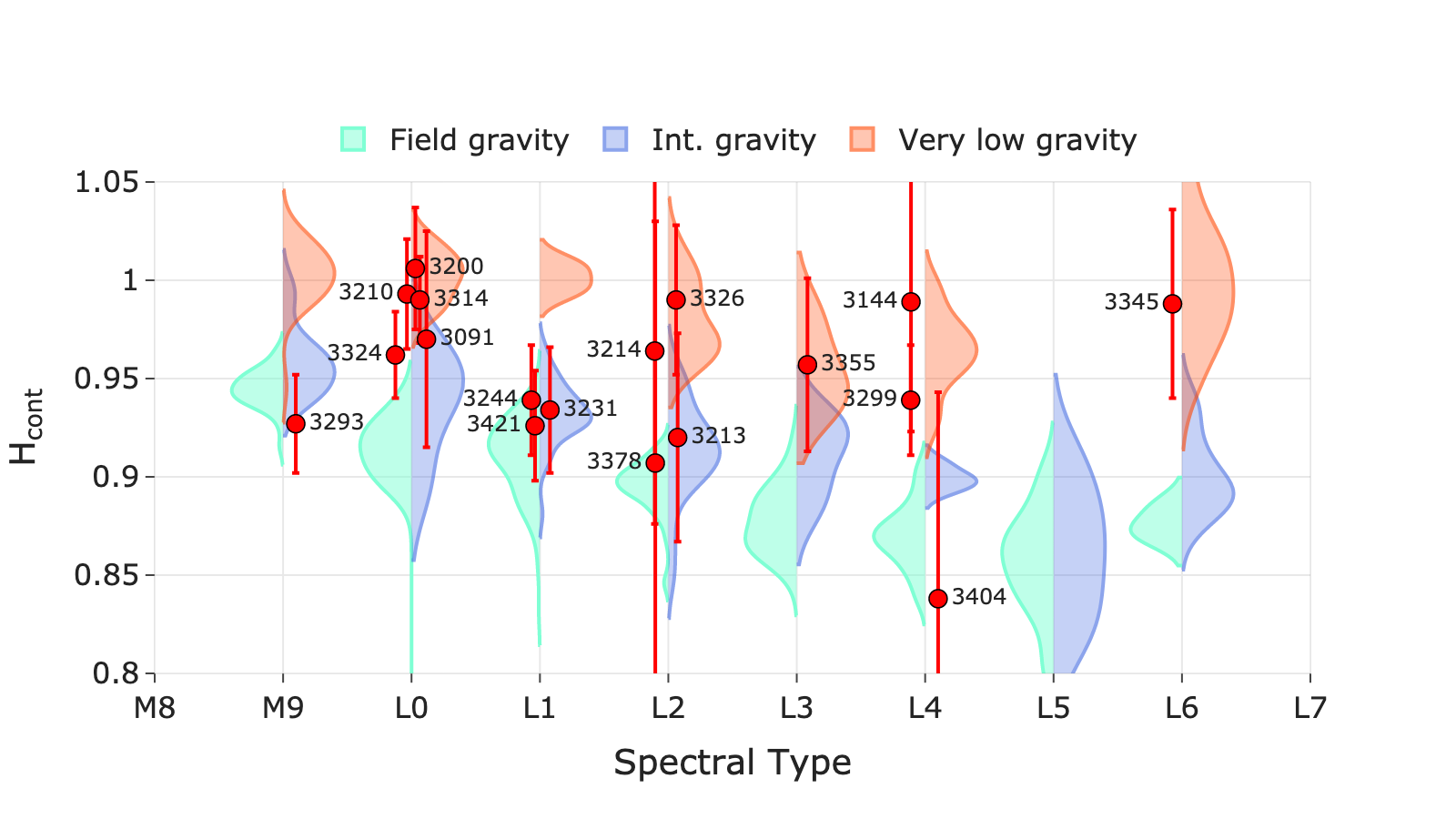}
      \caption{H$_{\rm cont}$ gravity index from \citet{Allers2013} for our targets (red dots) over-plotted over a violin graph of the distributions for ultracool dwarfs with very-low gravity (red), intermediate gravity (blue) and field-gravity (cyan) from the SPEX library of ultracool dwarfs. Our targets are shifted randomly horizontally for clarity.}
         \label{fig:gravity}
\end{figure*}

 We measured the H$_{\rm cont}$ index in all our spectra, as well as in  the 891 spectra from the SPEX Ultracool dwarfs library with a good quality (\texttt{QUALITY\_FLAG=OK} and \texttt{MEDIAN\_SNR}$\ge$50) and a resolution R$\ge$120. The uncertainty was estimated by simply propagating the standard errors of the means used in the H$_{\rm cont}$ index formula. We derived near-infrared spectral types for each spectrum from the SPEX Ultracool dwarfs library based on the \citet{Kirkpatrick2010} L dwarf classification scheme and a gravity classification between very-low (\texttt{VL-G}), intermediate (\texttt{INT-G}) and field (\texttt{FLD-G}) gravity using \citet{Allers2013} classification scheme. Figure~\ref{fig:gravity} shows the results in the form of a violin graph, using the spectral types that will be presented in Section~\ref{sec:spt} for our targets. The index measurement, spectral type and gravity classification for the 891 spectra are available in electronic form in Table~\ref{tab_all_indices}.
 
 Within the relatively large error bars, a number of our targets seem to have an H$_{\rm cont}$ favoring low or intermediate gravity and hence a young age. These include 3210, 3200, 3314, 3091, 3326, 3214, 3355, 3144, 3299, 3345. Two objects have H$_{\rm cont}$ values more consistent with high-gravity older objects: 3404 and 3293, although the large uncertainties make them also consistent with intermediate gravity objects. Objects 3244, 3421, 3231 have H$_{\rm cont}$ values consistent with intermediate or low gravity, while 3378 and 3213 have inconclusive H$_{\rm cont}$ indices compatible within the uncertainties either with high or intermediate gravity objects.

\begin{figure*}
   \centering
   \includegraphics[width=0.95\textwidth]{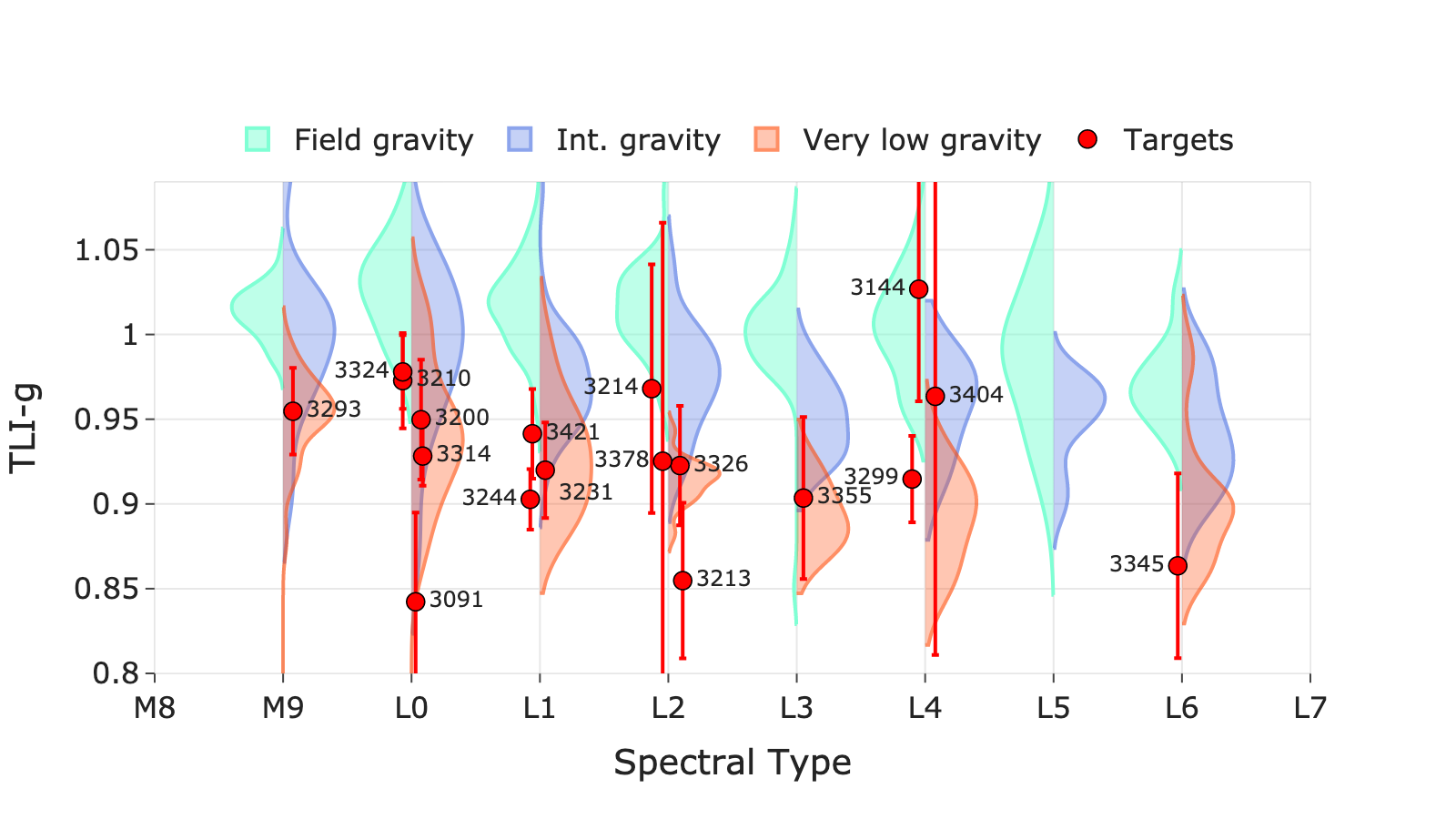}
      \caption{TLI-g gravity index from \citet{Almendros2022} for our targets (red dots) over-plotted over a violin graph of the distributions for ultracool dwarfs with very-low gravity (red), intermediate gravity (blue) and field-gravity (cyan) from the SPEX library of ultracool dwarfs. Our targets are shifted randomly horizontally for clarity.}
         \label{fig:gravityTLIg}
\end{figure*}

\subsection{TLI-g index}
The TLI-g index was invented recently by \citet{Almendros2022} using machine learning techniques to specifically distinguish low and field gravity ultracool objects. We measured the TLI-g index in all our spectra, as well as in  the 891 spectra from the SPEX library mentioned in the previous section (see Table~\ref{tab_all_indices}). Figure~\ref{fig:gravityTLIg} shows the results in the form of a violin graph.

Within the error bars we can see that all our targets have a TLI-g index favoring low or intermediate gravity and hence a young age except in the case of:
\begin{itemize}
\item 3144 with a TLI-g index favoring a field gravity and hence an older age
\item 3378 and 3404 have such large uncertainties that the TLI-g index is inconclusive
\end{itemize}
Although \citet{Almendros2022} defined the TLI-g index using a sample of M0--L3 ultracool dwarfs, we can see in Fig.~\ref{fig:gravityTLIg} that it seems to work equally well on the L4 and L6 dwarfs of our sample.

\subsection{Final youth status}
 Table~\ref{tab_youth} gives a summary of the 4 youth diagnostics as well as the final status for each target. Among the 17 targets, 9 have the four diagnostics indicating a young age, and are therefore firmly confirmed as young ultracool dwarfs members of the USco and Oph associations. Another 5 have inconclusive H$_{\rm cont}$ indices within the large uncertainties and/or inconclusive comparison with standards but have all other diagnostics indicating low-gravity, and are therefore confirmed as young ultracool dwarfs with a high level of confidence as well. 
 
 The remaining cases are discussed here:
 \begin{itemize}
 \item 3144 is classified as young from the $J-Ks$ color, the comparison with spectral standards, and the H$_{\rm cont}$ index, but as old from the TLI-g index. Extinction can affect the TLI-g measurement, but would move the object down in the diagram and make it look younger. Instead we find that 3144 has a higher TLI-g value. Its value is still consistent with intermediate gravity object within the uncertainties and given that all the other diagnostics are favoring a young age we classify 3144 as a young ultracool dwarf as well.
 \item 3404 is classified as young using 2 diagnostics and is only classified as possibly old using the H$_{\rm cont}$ index while the TLI-g index is inconclusive because of the large uncertainties. Given that the large uncertainties on the  H$_{\rm cont}$ index make it fully compatible with intermediate and low gravity objects as well, we classify 3404 as young as well.
 \item 3293  has a marginally better fit with an old standard and has a H$_{\rm cont}$ index clearly favoring a high-gravity older object. But the $J-Ks$ color and TLI-g index are favoring a young object. The integrated line-of-sight extinction at 250~pc is relatively large (A$_{\rm V}$=2.99~mag) and its $J-Ks$ color and TLI-g index might therefore be affected by reddening. With the current data it is difficult to draw any conclusion on 3293 youth.
 \end{itemize}
 
 In conclusion, 16 of the 17 targets have multiple evidences of youth and one (3293) is inconclusive. In total we therefore firmly confirm the youth of 16 candidates out of 17 as young L-dwarfs members of the USco and Oph associations, or 94\%.

\begin{table}
\centering

\caption{Diagnostics of youth  \label{tab_youth}}
\begin{tabular}{cccccc}

\hline
  ID &   $J-Ks$ & Comparison    &     H$_{\rm cont}$ & TLI-g & Young?   \\
     &          & with standard &                    &       &     \\
\hline
  3091 & Young & Young  & Young & Young & Young\\
  3144 & Young & Young  & Young & Old   & Young\\
  3200 & Young & Young  & Young & Young & Young\\
  3210 & Young & Young  & Young & Young & Young\\
  3213 & Young & Young  & ?     & Young & Young\\
  3214 & Young & Young  & Young & Young & Young\\
  3231 & Young & Young? & Int?  & Young & Young\\
  3244 & Young & Young? & Int?  & Young & Young\\
  3293 & Young & Old?   & Old?  & Young & ?    \\
  3299 & Young & Young  & Young & Young & Young\\
  3314 & Young & Young  & Young & Young & Young\\
  3326 & Young & Young  & Young & Young & Young\\
  3345 & Young & Young  & Young & Young & Young\\
  3355 & Young & Young  & Young & Young & Young\\
  3378 & Young & Young  & ?     & ?     & Young\\
  3404 & Young & Young? & Old?  & ?     & Young\\
  3421 & Young & Young? & Int?  & Young & Young\\
\hline
\end{tabular}
\end{table}

\section{Spectral types, effective temperature and masses \label{sec:spt}}

\begin{figure*}
   \centering
   \includegraphics[width=0.9\textwidth]{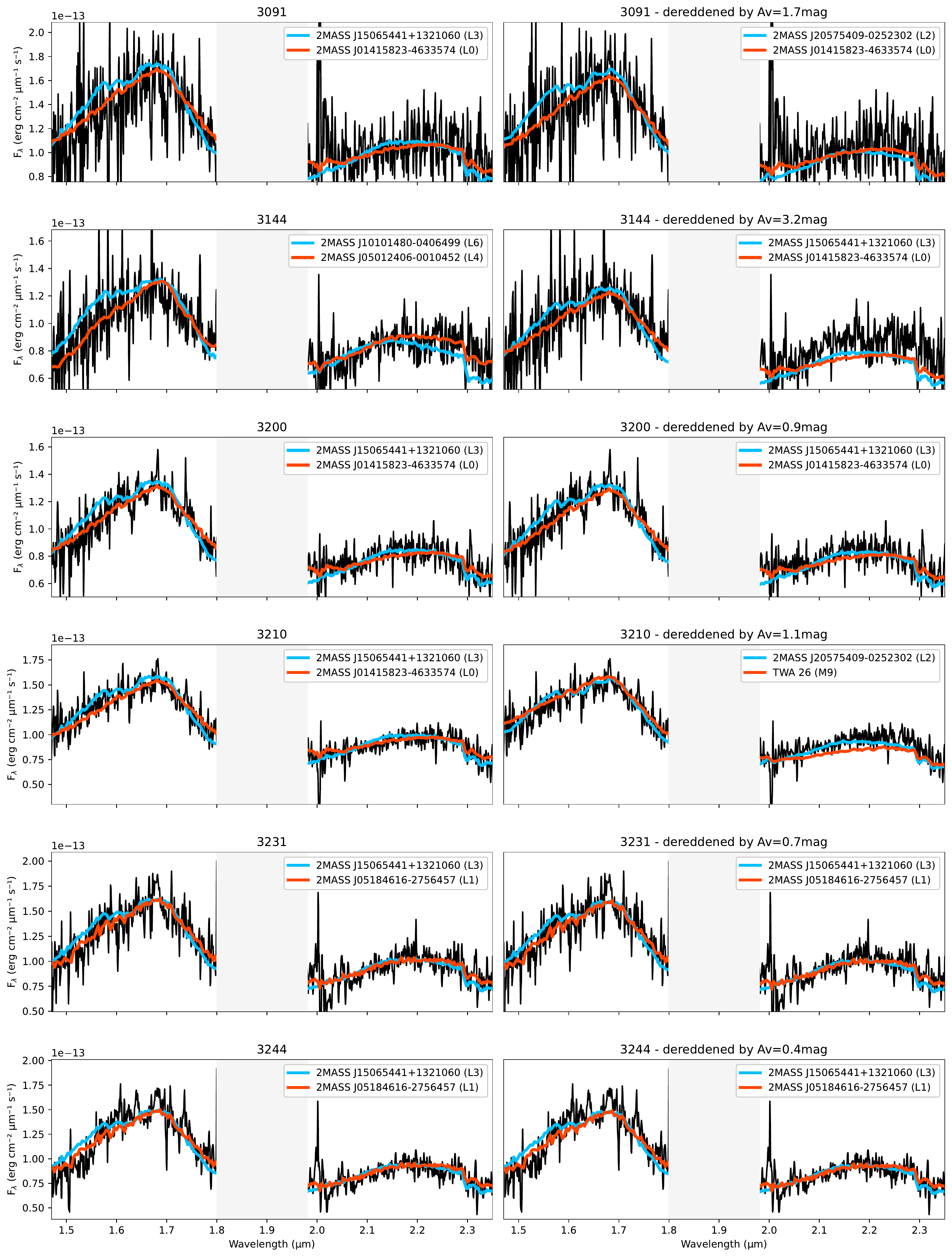}
      \caption{Comparison of our targets spectra (black) with very-low gravity standards (red) and field-gravity standards (blue). The left panels show the original spectrum, and the right panels show the spectrum derredened by the cumulative line-of-sight extinction indicated in the plot title and in Table~\ref{tab_targets}}
         \label{fig:spt1}
\end{figure*}

\begin{figure*}
   \centering
   \includegraphics[width=0.9\textwidth]{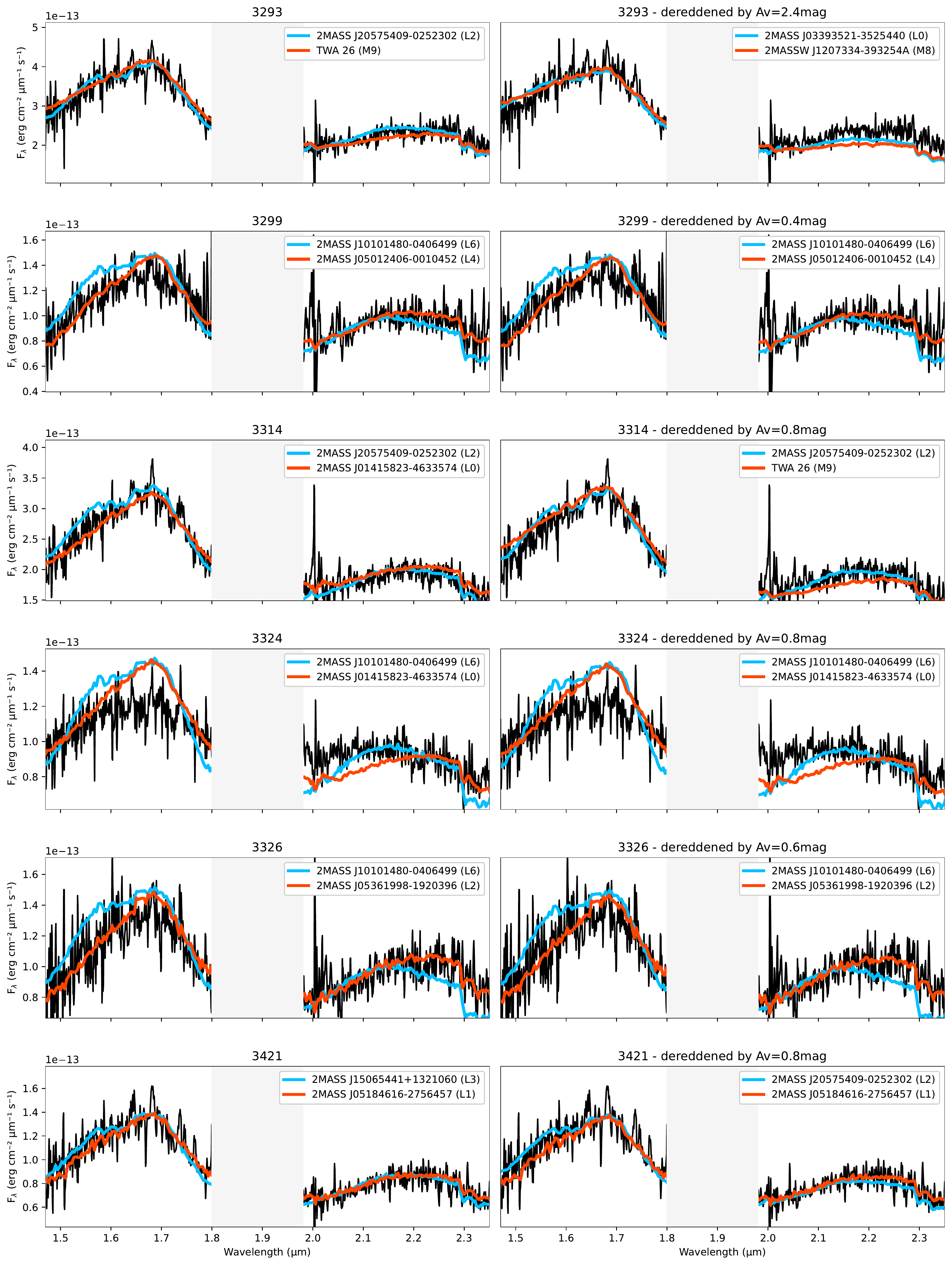}
      \caption{Same as Fig.~\ref{fig:spt1}}
         \label{fig:spt2}
\end{figure*}

\begin{figure*}
   \centering
   \includegraphics[width=0.9\textwidth]{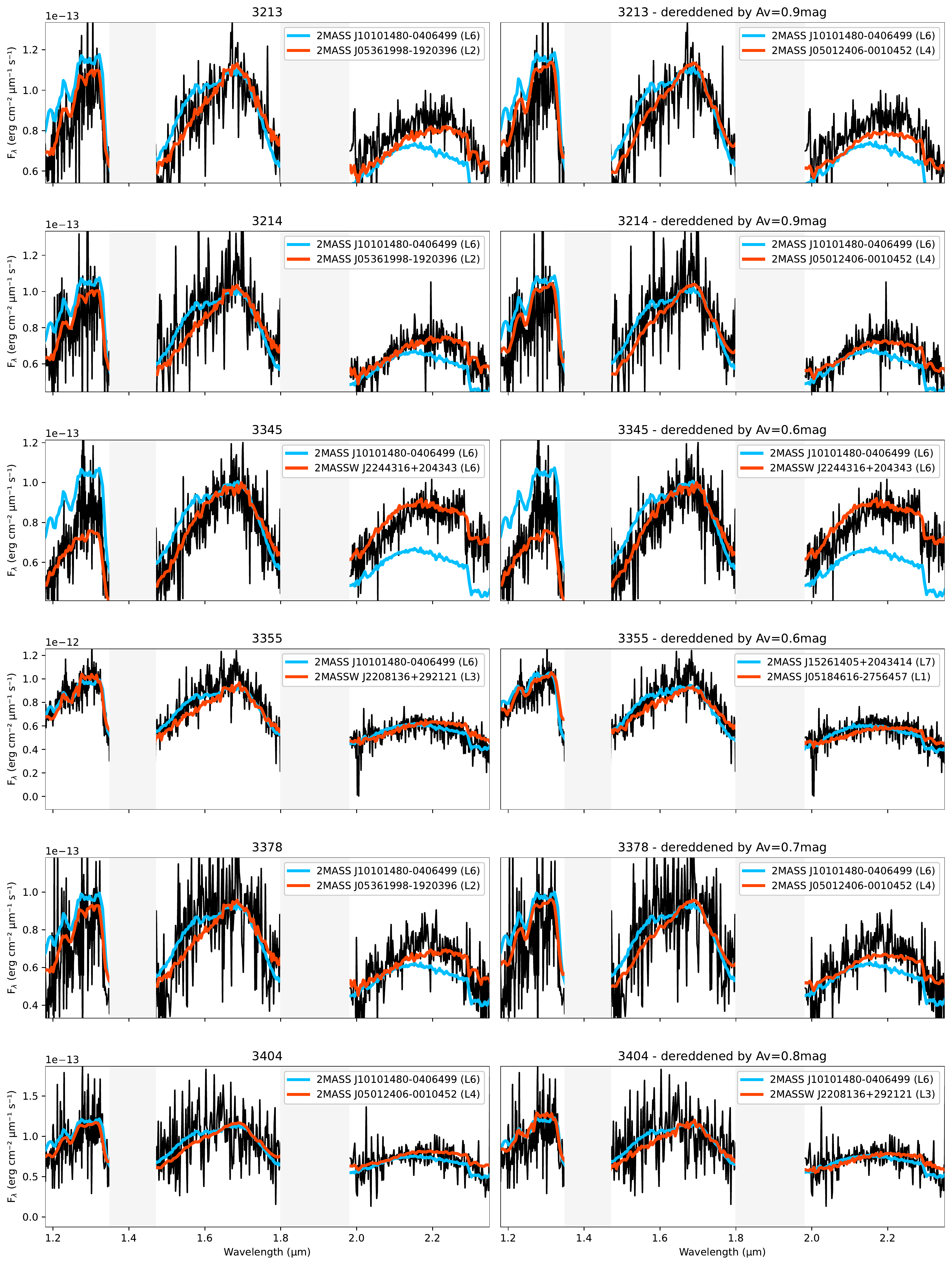}
      \caption{Same as Fig.~\ref{fig:spt1}}
         \label{fig:spt3}
\end{figure*}

We now use Figures~\ref{fig:spt1} to \ref{fig:spt3} to estimate the spectral types of the 16 targets with evidences of youth identified in the previous section. With the question of their youth settled, we are left to decide whether extinction affects the spectrum of the targets. By using the integrated line-of-sight extinction towards each target until 250~pc, we can partially break the degeneracy between spectral type and reddening and check the worst-case scenario in which the target eventually lies at the far edge of the association.

Figures~\ref{fig:spt1} to \ref{fig:spt3} show that extinction does not affect the result of the comparison for 3091, 3200, 3231, 3244, 3299, 3324, 3326, 3421, 3345. All indeed have fairly small integrated extinction (see Table~\ref{tab_targets}).

The match is significantly better without extinction in the cases of 3213 (L2), 3314 (L0), 3210 (L0), 3144 (L4), specially in the K-band. 

Finally, the match is equally good with or without dereddening for 3404 (L3--L4), 3378 (L2--L4), 3355 (L1--L3), 3214 (L2--L4). We note that the difference is always smaller than 2 subclasses, and we adopt the corresponding ranges as final spectral type.

These results show that reddening does not affect much our target, in agreement with the integrated line-of-sight extinction at 250~pc, and confirm that they are most likely not reddened background contaminants.

\begin{figure}
   \centering
   \includegraphics[width=0.45\textwidth]{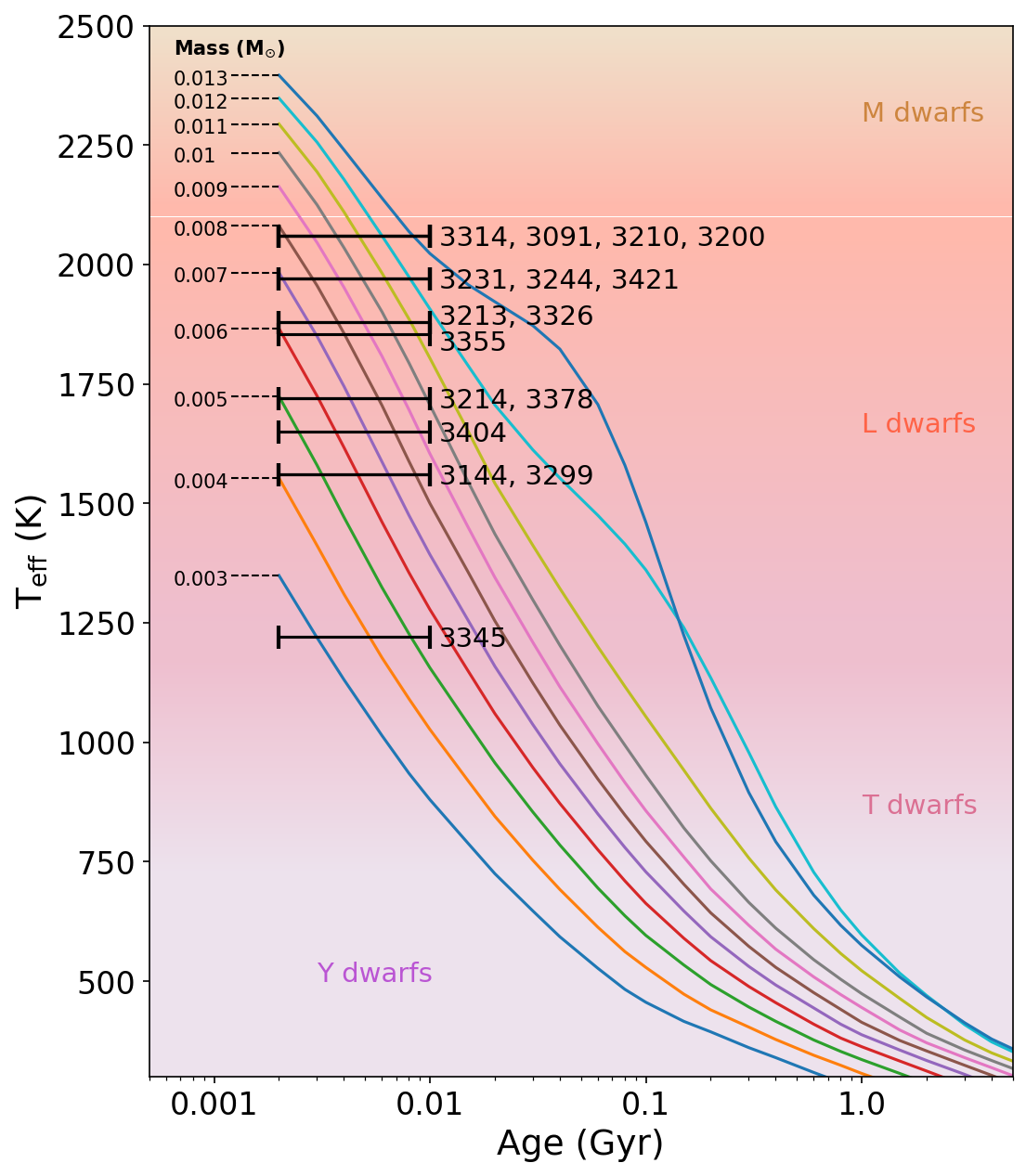}
      \caption{Effective temperature vs age between 1~Myr and 15~Gyr according to the \citet{Marley2021} evolutionary models. The targets are represented as well assuming an age between 1 and 10~Myr.
         \label{fig:evolution}}
\end{figure}

   These spectral types are translated into effective temperatures using the empirical relationship for young L-dwarfs reported in \citet{Faherty2016}, which in turn are translated into masses using the  \citet{Saumon2008} models for 3, 6 and 10~Myr. Given their location in the Scorpius OB2 complex, most of our targets are expected to belong to Upper Scorpius and have ages in the range 6--10~Myr. Three of them (3421, 3293, 3144, see Fig.~\ref{fig:map}) are nevertheless located on top of the $\rho-$Ophiuchus molecular clouds and probably belong to the young (1$\sim$3Myr) association. Uncertainties on the spectral types (1 to 2 subclasses) translate into uncertainties of the order of 250~K for the effective temperatures and 0.002~M$_{\odot}$ for the masses at a given age. Table~\ref{tab_spt} gives the results and shows that all the candidates seem to have masses in the planetary domain, the least massive having a mass of only 0.004--0.006~M$_{\odot}$ depending on the age. These objects will cool steadily over time, dissolve in the galactic field population and become field T and Y-dwarfs within the next 50$\sim$100~Myr, as illustrated in Figure~\ref{fig:evolution}.

\begin{table}
\centering
\caption{Adopted spectral types and estimated effective temperatures and masses for the confirmed members. \label{tab_spt}}
\scriptsize
\begin{tabular}{cccccc}
\hline
  ID &   SpT & T$_{\rm eff}$     & Mass 3Myr & Mass 6Myr & Mass 10Myr \\
     &       &  (K)              & (M$_{\odot}$) & (M$_{\odot}$) & (M$_{\odot}$)\\
\hline
3213 & L2 & 1880                          & 0.007     & 0.010 & 0.012 \\
3214 & L2--L4 & 1560--1880      & 0.005--0.007       & 0.007--0.010 & 0.009--0.012 \\
3345 & L6 & 1220                          & 0.003     & 0.004& 0.006 \\
3355 & L1--L3 & 1740--1970                & 0.006--0.008     & 0.008--0.011& 0.011--0.013 \\
3378 & L2--L4 & 1560--1880                & 0.005--0.007     & 0.007--0.010 & 0.009--0.012 \\
3404 & L3--L4 & 1560--1740                & 0.005--0.006     & 0.007--0.008 & 0.009--0.011 \\
3314 & L0 & 2060 		          & 0.009     & 0.012& 0.014 \\
3231 & L1 & 1970  		          & 0.008     & 0.011& 0.013 \\
3091 & L0 & 2060   	                  & 0.009     & 0.012& 0.014 \\
3210 & L0 & 2060  		          & 0.009     & 0.012& 0.014 \\
3244 & L1 & 1970 		          & 0.008     & 0.011& 0.013 \\
3421 & L1 & 1970 		          & 0.008     & 0.011& 0.013 \\
3326 & L2 & 1880  		          & 0.007     & 0.010 & 0.012 \\
3144 & L4 & 1560 		          & 0.005     & 0.007& 0.009 \\
3200 & L0 & 2060  		          & 0.009     & 0.012& 0.014 \\
3299 & L4 & 1560                          & 0.005     & 0.007& 0.009 \\
\hline
\end{tabular}
\end{table}

\section{Objects of interest}
In the following we discuss a couple of objects of interest: the coolest of our targets and a ultra-wide pair of planetary mass objects.

\subsection{The coolest target: DANCe J16081299-2304316 (3345)}
In this section we compare the spectrum of the coolest of our targets, the L6 DANCe J16081299-2304316 (3345), with spectra of exoplanets and free-floating planets with similar ages and spectral types from the literature.

The H-band is in general well matched by all these young exoplanet and free-floating planet spectra, but the overall slope of DANCe J16081299-2304316 is shallower than that of all these objects and the higher J-band flux must be due to the slightly earlier spectral type. The 2.3$\mu$m CO overtone is well matched in most free-floating planet spectra but is not as pronounced in the exoplanet spectra. On the other hand the drop observed in DANCe J16081299-2304316 at wavelength greater than 2.3$\mu$m is observed in HR8799c spectrum only, making it a good free-floating analog of this directly imaged young gas-giant planet. Overall, the spectrum of DANCe J16081299-2304316 appears most similar to that of HR8799c, and both objects have near-infrared spectral types of L6. 

\begin{figure*}
   \centering
   \includegraphics[width=0.95\textwidth]{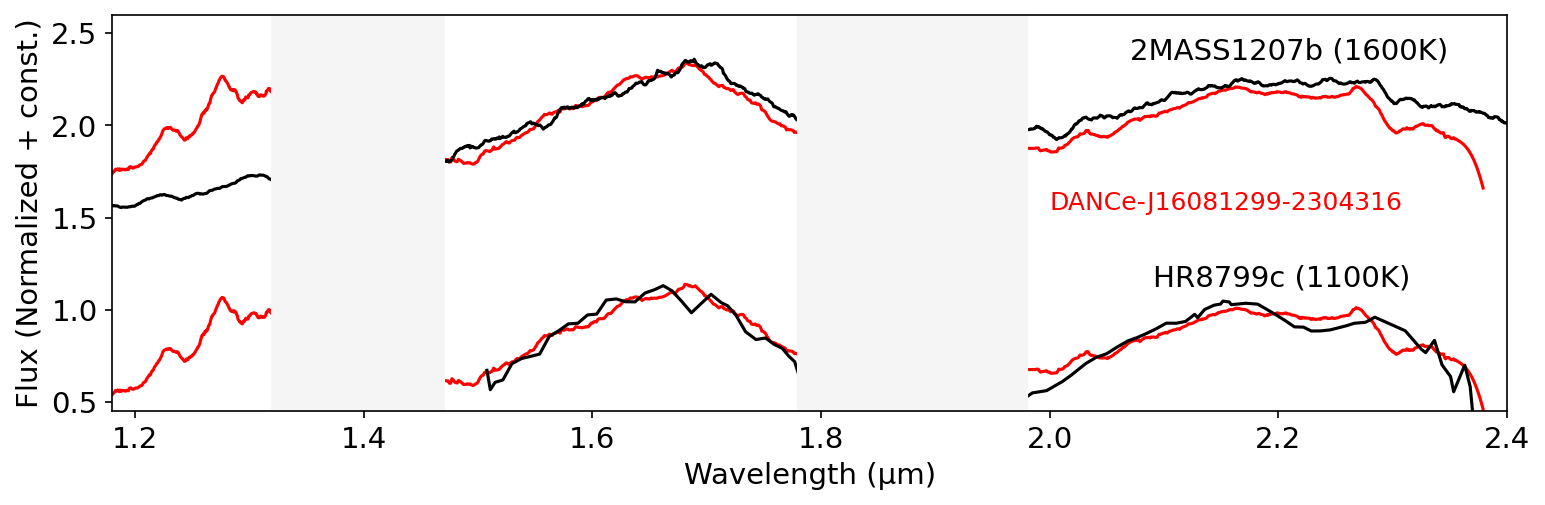}
      \caption{Smoothed spectrum of DANCe J16081299-2304316 (red) and the planetary mass companion 2MASS J12073346-3932539 b  and HR8799c from \citet{Greenbaum2018}.
         \label{fig:comp_planets}}
\end{figure*}

\begin{figure*}
   \centering
   \includegraphics[width=0.95\textwidth]{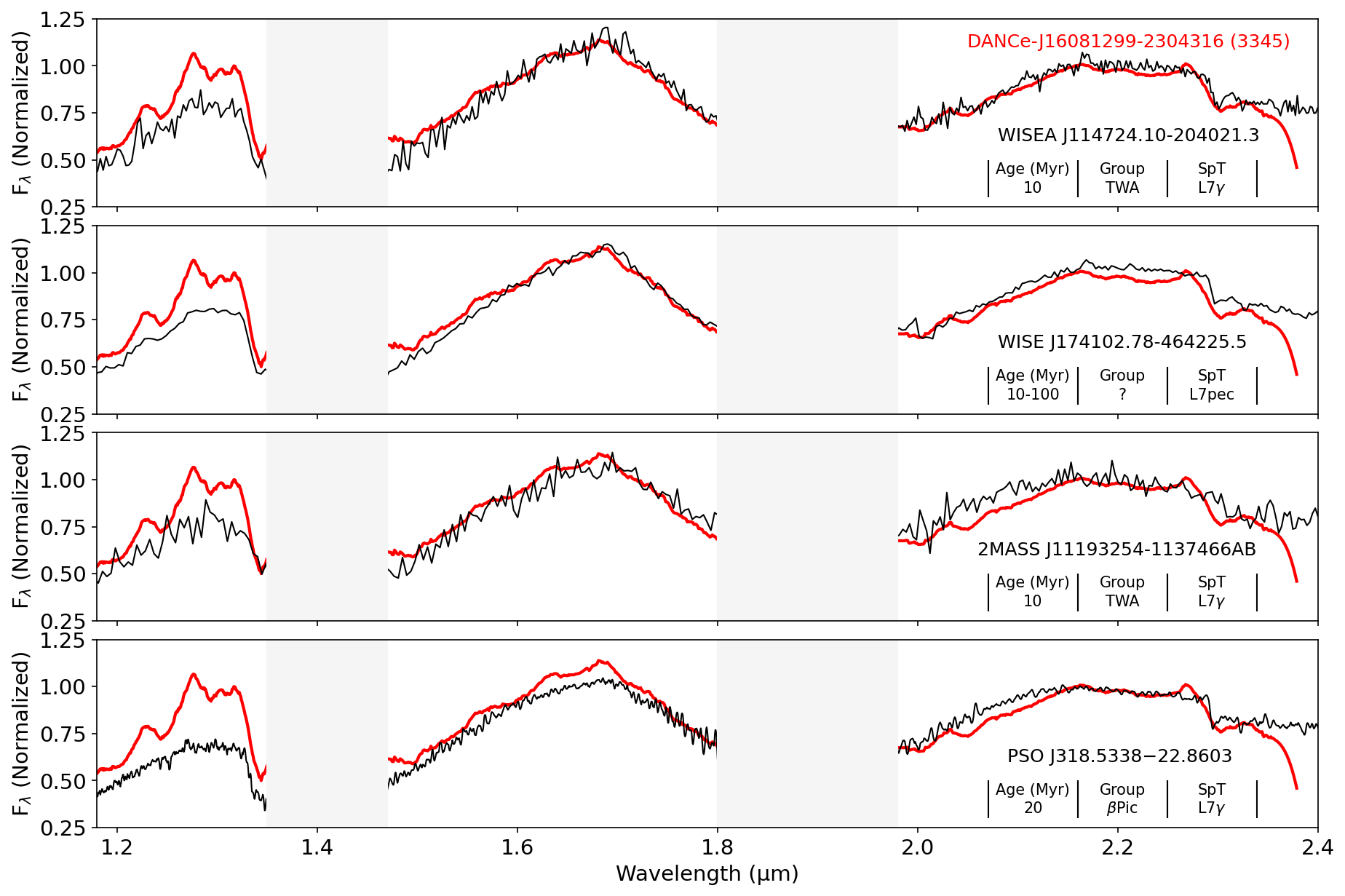}
      \caption{Smoothed spectrum of DANCe J16081299-2304316 (red) compared to the TW Hydra L7$\gamma$ free-floating planets WISEA J114724.10-204021.3 \citep{Schneider2016} and 2MASS J11193254-1137466AB \citep{Kellogg2016}, the young L7pec free-floating planet WISE J174102.78-464225.5 \citep{Schneider2014}, and the $\beta$-Pic L7$\gamma$ free-floating planet PSO J318.5338−22.8603 \citep{Liu2013} }
         \label{fig:comp_ffp}
\end{figure*}

\subsection{DANCe J16135217-2443562  and DANCe J16134589-2442310: an ultra-wide pair?} 
\begin{figure}
   \centering
   \includegraphics[width=0.45\textwidth]{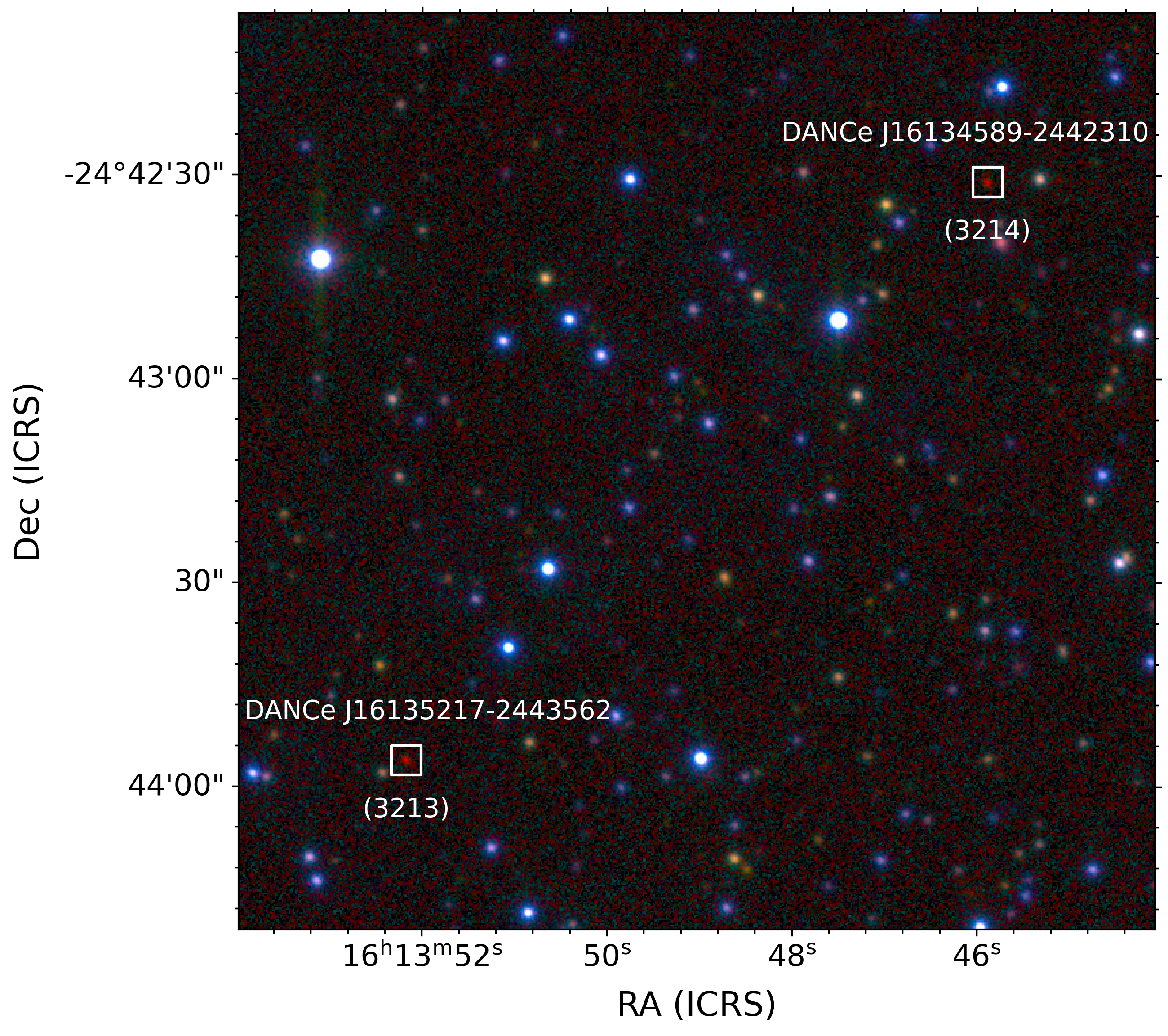}
      \caption{Three color image ($r,Y,Ks$ as blue, green, red) of the field around DANCe J16135217-2443562 and DANCe J16134589-2442310. Both are indicated by a square.}
         \label{fig:bin}
\end{figure}

DANCe J16135217-2443562 (3213) and DANCe J16134589-2442310 (3214) form a wide visual pair with a separation on 120\arcsec corresponding to a projected separation of $\sim$17\,400~A.U. at a distance of 145pc, as shown in Fig.~\ref{fig:bin}. Such a wide separation for such low mass objects suggest that it is probably a coincidence rather than a bound physical pair. But on the other hand the very low spatial density of free-floating planets reported in \citet{2022NatAs...6...89M} (between 0.4 and 1 FFP per square degree) suggests that such a coincidence is highly unlikely, and calls for follow-up observations of this intriguing pair. Improved proper motions measurements, as well as parallaxes and radial velocities would help understand and eventually confirm their common origin. Such a pair could indeed also be a remnant of an extreme case of ultra-wide multiple system such as the ones reported in Taurus \citep{Joncour2017} or the result of a simultaneous dynamical ejection of two planets in a planetary system.

\section{Conclusions}

We have obtained near-infrared spectra of 18 ultracool candidate members of Upper Scorpius and Ophiuchus discovered by \citet{2022NatAs...6...89M} using SWIMS at the Subaru telescope and EMIR at the Grantecan telescope. One of the spectra was affected by the poor ambient conditions (clouds) and we discard it in the analysis. 

The spectra allow us to confirm the low gravity, and hence youth, using 4 diagnostics: i) the shape of their H-band continuum measured by the H$_{\rm cont}$ index; ii) their $J-Ks$ color redder than field counterparts; iii) by comparison with near-infrared spectra of young L-dwarf standards; iv) using the TLI-g gravity sensitive index. Among the 17 targets, 16 have multiple evidences of youth and one (3293) is inconclusive. In total we therefore firmly confirm the youth of 16 candidates out of 17 as young L-dwarfs members of the USco or Ophiuchus association, corresponding to a contamination rate of only $\lesssim$6\% and indicating that the methodology devised by \citet{Bouy2013} and \citet{Sarro2014} and used by \citet{2022NatAs...6...89M} is very reliable. 

 The spectral types of the targets are estimated by comparison with young L-dwarf standards, and range between L0 and L6. Using \citet{Faherty2016} empirical relationship for young L-dwarfs we transform these spectral types into effective temperatures, and found that the objects have temperatures in the range between 1\,220 and 2\,060~K, corresponding to masses in the range 0.004--0.013~M$_{\odot}$ according to the models of \citet{Saumon2008} for ages between 3 and 10~Myr, consistent with \citet{2022NatAs...6...89M} estimate based on the photometry only. Interestingly even the brightest target (DANCe J16064553-2121595 = 3355) is an early L-dwarf, suggesting that many objects fainter than $M_{\rm J}\gtrsim10.5$~mag must indeed have masses in the planetary mass domain.

\begin{acknowledgements}
We thank A. Burgasser for his help with \emph{SPLAT}. We are grateful to M. Bonnefoy, M. Liu, K. Luhman and B. Bowler for sharing spectra of young L-dwarfs and planets. This research has received funding from the European Research Council (ERC) under the European Union’s Horizon 2020 research and innovation programme (grant agreement No 682903, P.I. H. Bouy), and from the French State in the framework of the ”Investments for the future” Program, IdEx Bordeaux, reference ANR-10-IDEX-03-02. P.A.B. Galli acknowledges financial support from São Paulo Research Foundation (FAPESP) under grants 2020/12518-8 and 2021/11778-9. DB and NH have been partially funded by the Spanish State Research Agency (AEI) Project No. PID2019-107061GB-C61 and MDM-2017-0737 Unidad de Excelencia {\em Mar\'{\i}a de Maeztu} - Centro de Astrobiolog\'{\i}a (CSIC-INTA).
Based on observations made with the Gran Telescopio Canarias (GTC), installed in the Spanish Observatorio del Roque de los Muchachos of the Instituto de Astrofísica de Canarias, in the island of La Palma. This work is partly based on data obtained with the instrument EMIR, built by a Consortium led by the Instituto de Astrofísica de Canarias. EMIR was funded by GRANTECAN and the National Plan of Astronomy and Astrophysics of the Spanish Government.
Based in part on data collected at Subaru Telescope which is operated by the National Astronomical Observatory of Japan and obtained from the SMOKA, which is operated by the Astronomy Data Center, National Astronomical Observatory of Japan. 
This research has made use of the VizieR catalogue access tool, CDS, Strasbourg, France. The original description of the VizieR service was published in A\&AS 143, 23.
This work made use of GNU Parallel \citep{Tange2011a}, astropy \citep{astropy2013,astropy2018}, Topcat \citep{Topcat}, specutils \citep{specutils}, matplotlib \citep{matplotlib}, Plotly \citep{plotly}, Numpy \citep{numpy}, aplpy \citep{aplpy2012,aplpy2019}.
\end{acknowledgements}

\bibliographystyle{aa} 

\begin{appendix}
\section{Tables}

\begin{table}
\centering
\caption{Gravity sensitive indices \label{tab_indices}}
\begin{tabular}{ccc}
\hline
Object & H$_{\rm cont}$ & TLI-g \\
\hline
  3091 & 0.97$\pm$0.05 & 0.84$\pm$0.05\\
  3144 & 0.99$\pm$0.07 & 1.03$\pm$0.07\\
  3200 & 1.01$\pm$0.03 & 0.95$\pm$0.04\\
  3210 & 0.99$\pm$0.03 & 0.97$\pm$0.03\\
  3213 & 0.92$\pm$0.05 & 0.85$\pm$0.05\\
  3214 & 0.96$\pm$0.09 & 0.97$\pm$0.07\\
  3231 & 0.93$\pm$0.03 & 0.92$\pm$0.03\\
  3244 & 0.94$\pm$0.03 & 0.90$\pm$0.02\\
  3293 & 0.93$\pm$0.03 & 0.95$\pm$0.03\\
  3299 & 0.94$\pm$0.03 & 0.91$\pm$0.03\\
  3314 & 0.99$\pm$0.02 & 0.93$\pm$0.02\\
  3324 & 0.96$\pm$0.02 & 0.98$\pm$0.02\\
  3326 & 0.99$\pm$0.04 & 0.92$\pm$0.04\\
  3345 & 0.99$\pm$0.05 & 0.86$\pm$0.05\\
  3355 & 0.96$\pm$0.04 & 0.90$\pm$0.05\\
  3378 & 0.91$\pm$0.12 & 0.93$\pm$0.14\\
  3404 & 0.84$\pm$0.10 & 0.96$\pm$0.15\\
  3421 & 0.93$\pm$0.03 & 0.94$\pm$0.03\\
\hline\end{tabular}
\end{table}

\begin{table*}
\caption{SPEX library of very-low gravity ultracool standards \label{splat_stds}}
\begin{tabular}{lcccc}
\hline
  Name & RA & Dec & SpT & Ref. \\
\hline
  2MASSW J1207334-393254A & 181.88959 & -39.548443 & M8 & (1) \\
  TWA 26 & 174.96304 & -31.989279 & M9 & (1) \\
  2MASS J01415823-4633574 & 25.492626 & -46.565945 & L0$\gamma$ & (2)\\
  2MASS J05184616-2756457 & 79.692337 & -27.946028 & L1$\gamma$ & (3)\\
  2MASS J05361998-1920396 & 84.083244 & -19.344334 & L2$\gamma$ & (3)\\
  2MASSW J2208136+292121 & 332.05679 & 29.355972 & L3$\gamma$ & (2)\\
  2MASS J05012406-0010452 & 75.35025 & -0.17922223 & L4$\gamma$ & (2)\\
  2MASSW J2244316+204343 & 341.13196 & 20.728695 & L7$\gamma$ & (4) \\
\hline\end{tabular}
\tablebib{(1) \citet{Gizis2002} ; (2) \citet{Cruz2009} ; (3) \citet{Bardalez2014} ; (4) \citet{Kirkpatrick2008}}
\end{table*}

\begin{table*}
\caption{Library of field ultracool standards \label{old_stds}}
\begin{tabular}{lcccc}
\hline
  Name &  RA (J2000) &  Dec (J2000) &  SpT &  Ref. \\
\hline
  VB 10 & 289.2400917 & 5.1506056 & M8 & (2)\\
  LP 944-20 & 54.89675 & -35.4289139 & L0 & (1) \\
  GJ 1048B & 38.9997083 & -23.5223611 & L1 & (1)\\
  2MASSI J2057540-025230 & 314.4753917 & -2.8750722 & L2 & (2)\\
  2MASSW J1506544+132106 & 226.7267125 & 13.3516889 & L3 & (3)\\
  2MASSW J0036159+182110 & 9.0674 & 18.3529083 & L4 & (1) \\
  SDSSp J144600.60+002452.0 & 221.5025833 & 0.4144444 & L5 & (4) \\
  2MASSI J1010148-040649 & 152.5616958 & -4.1138694 & L6 & (5) \\
  2MASSI J1526140+204341 & 231.5585417 & 20.7281833 & L7 & (2)\\
\hline\end{tabular}
\tablebib{(1) \citet{Burgasser2008} ; (2) \citet{Burgasser2004} ; (3) \citet{Burgasser2007} ; (4) \citet{Bardalez2014} ; (5) \citet{Reid2006} }
\end{table*}

\begin{table}
\centering
\caption{Gravity sensitive indices measured in 891 spectra of the SPEX library \label{tab_all_indices}}
\begin{tabular}{lcccccccc}
\hline
Name & RA (J2000) & Dec (J2000) & SpT & \texttt{SPEX\_CLASS} & H$_{\rm cont}$ &  e\_H$_{\rm cont}$ & TLI-g & e\_TLI-g \\
\hline
    2MASS J0000286-124515 & 0.11945834 & -12.75425 & M9.0 & FLD-G & 0.947 & 0.003 & 1.015 & 0.01\\
  2MASS J00013044+1010146 & 0.37683335 & 10.170723 & M6.0 & VL-G & 0.988 & 0.006 & 1.009 & 0.023\\
  LHS 102B & 1.1451666 & -40.734943 & L4.0 & FLD-G & 0.855 & 0.002 & 1.011 & 0.008\\
  LEHPM 1-162 & 1.4486667 & -21.954889 & M8.0 & FLD-G & 0.963 & 0.002 & 0.994 & 0.009\\
  2MASSI J0006205-172051 & 1.5854167 & -17.347389 & L2.0 & FLD-G & 0.897 & 0.006 & 0.998 & 0.006\\
  SDSS J000632.60+140606.4 & 1.6358334 & 14.101778 & L1.0 & FLD-G & 0.931 & 0.009 & 0.989 & 0.017\\
  SDSS J000649.16-085246.3 & 1.7048334 & -8.879528 & M8.0 & FLD-G & 0.921 & 0.004 & 1.035 & 0.008\\
  2MASS J0007078-245804 & 1.7827917 & -24.967834 & M8.0 & INT-G & 0.987 & 0.003 & 1.008 & 0.017\\
  2MASS J00085931+2911521 & 2.2471251 & 29.197805 & M8.0 & INT-G & 0.967 & 0.002 & 1.013 & 0.007\\
  2MASS J00100009-2031122 & 2.500375 & -20.520056 & M8.0 & FLD-G & 0.933 & 0.003 & 1.049 & 0.003\\
  2MASSI J0013578-223520 & 3.4907918 & -22.58889 & L5.0 & FLD-G & 0.859 & 0.01 & 0.992 & 0.01\\
  WISE J001450.14-083823.1 & 3.7089167 & -8.6397219 & M7.0 & FLD-G & 0.945 & 0.008 & 1.042 & 0.016\\
  2MASS J00145575-4844171 & 3.7322917 & -48.738083 & L3.0 & FLD-G & 0.883 & 0.006 & 1.022 & 0.013\\
  2MASSW J0015447+351603 & 3.9364998 & 35.267387 & L2.0 & INT-G & 0.9 & 0.003 & 1.025 & 0.006\\
  SDSS J001608.44-004302.3 & 4.0351253 & -0.71722221 & L3.0 & INT-G & 0.898 & 0.008 & 0.961 & 0.023\\
  2MASS J00163761+3448368 & 4.1567087 & 34.810223 & M8.0 & FLD-G & 0.939 & 0.004 & 1.016 & 0.013\\
\hline\end{tabular}
\end{table}

\end{appendix}

\end{document}